# Electrothermal Transport Induced Material Re-Configuration and Performance Degradation of CVD-Grown Monolayer MoS$_2$ Transistors

Ansh, Jeevesh Kumar, Gaurav Sheoran and Mayank Shrivastava

Department of Electronic Systems Engineering, Indian Institute of Science, Bengaluru, Karnataka, India, email: mayank@iisc.ac.in

**Abstract**

Device and material reliability of 2-dimensional materials, especially CVD-grown MoS$_2$, has remained un-addressed since 2011 when the first TMDC transistor was reported. For its potential application in next generation electronics, it is imperative to update our understanding of mechanisms through which MoS$_2$ transistors' performance degrades under long-term electrical stress. We report, for CVD-grown monolayer MoS$_2$, the very first results on temporal degradation of material and device performance under electrical stress. Both low and high field regimes of operation are explored at different temperatures, gate bias and stress cycles. During low field operation, current is found to saturate after hundreds of seconds of operation with the current decay time constant being a function of temperature and stress cycle. Current saturation after several seconds during low field operation occurs when a thermal equilibrium is established. However, high field operation, especially at low temperature, leads to impact ionization assisted material and device degradation. It is found that high field operation at low temperature results in amorphization of the channel and is verified by device and Kelvin Probe Force Microscopy (KPFM) analyses. In general, a prolonged room temperature operation of CVD-grown MoS$_2$ transistors lead to degraded gate control, higher OFF state current and negative shift in threshold voltage ($V_T$). This is further verified, through micro-Raman and Photoluminescence spectroscopy, which suggest that a steady state DC electrical stress leads to the formation of localized low resistance regions in the channel and a subsequent loss of transistor characteristics. Our findings unveil unique mechanism by which CVD MoS$_2$ undergoes material degradation under electrical stress and subsequent breakdown of transistor behavior. Such an understanding of material and device reliability helps in determining the safe operating regime from device as well as circuit perspective.

## I. Introduction

2-dimensional (2D) transition metal dichalcogenides (TMDCs), owing to their layered structure, offer much better immunity against scaling related challenges posed by Moore's law for transistors[1]. Besides, heavy electron effective mass in TMDCs strongly advocates their role as channel material for ultra-scaled field-effect transistors (FETs)[2] i.e. beyond 3 nm technology nodes. Looking at such promise as future device material, extensive research on FETs on various TMDCs has been conducted for more than a decade now. Several groups have reported few to single layer Molybdenum Disulfide (MoS$_2$) and other TMDC FETs[3-6]. Preliminary reports suggested remarkable ON/OFF current ratio[7], low sub-threshold slope (SS)[8] and large current density[5] - metrics important for acceptable transistor behavior for switching applications. However, large contact resistance ($R_C$) due to lack of covalent bonds[9] at the metal-MoS$_2$ interface has strongly questioned the ultimate deployment of MoS$_2$, and TMDCs in general, in the semiconductor industry. Several approaches like degenerate doping via surface charge transfer[10-13], dopant introduction (substitutional doping) during material growth[14], adding Chalcogen impurities at the contact[15] and selective phase transition[16] have been performed to get away with the $R_C$ bottleneck. Besides, TMDC heterostructures are being investigated for optoelectronic[17-19] and tunnel FET[20] applications. TMDC based memristors for neuromorphic applications[21-23] is another prospect that has evolved lately. Such technology driven reports are extensively available, however, very few reports discuss another crucial aspect – reliability. Apart from environment induced degradation of TMDCs[24, 25], very few reports elucidate the mechanism of electrical stress induced 2D material and device degradation[26-28].

Dattatray et. al. have reported the effect of oxygen and moisture on the transistor behavior especially in the hysteretic behavior[24]. In a separate effort, Bartolomeo et. al. have reported oxygen and moisture induced instabilities in the threshold voltage ($V_T$) of MoS$_2$ transistors upon multiple electrical measurements[25]. Unlike other reports on material-device degradation, Dominik et. al. reported the very first study on electrical breakdown of MoS$_2$ transistors with current densities 50 times higher than that of Copper[26]. Avalanche breakdown in exfoliated MoS$_2$ has been demonstrated and studied by Jinsu et. al[27] where critical field for avalanche multiplication in MoS$_2$ is calculated as a function of thickness of MoS$_2$ and temperature. Mishra et. al. reported electrical stress dependent material and device co-degradation study of Graphene transistors[28]. They elucidated the failure mechanism which Graphene transistors undergo in the presence of electrical stress and ambient oxygen. Stress induced temperature sets up favorable conditions for oxygen to react with graphene which proceeds in several steps until complete device failure is observed. Similar reports on electrical stress induced material and device performance degradation of TMDC FETs are unfortunately missing.

Fundamental properties like electrical and thermal conductivity play dominant role in determining the current carrying capacity and breakdown field for a material. These properties for MoS$_2$ are different from other semiconductors like Silicon and Graphene, essentially, because of difference in their molecular arrangements. Naturally, response of MoS$_2$ to electric field and high current densities must be different from that of other materials. Similarly, the mechanism through which degradation occurs in TMDC devices under electrical stress, in principle, must be different from those observed in Graphene or other semiconductors. In order to fabricate robust 2D TMDC transistors and mitigate reliability issues in these devices, to develop understanding of material and device degradation mechanisms is essential. Moreover, dynamic evolution of channel degradation and its impact on channel current can hinder critical electric field for acceptable device operation, especially in ultra-scaled transistors. As a result, reconsidering trends in voltage scaling for these materials may become imperative. Therefore, to understand the dynamics of material and device degradation in MoS$_2$ and subsequent failure of MoS$_2$ FETs, we start with observing steady state electrical behavior of Chemical Vapor Deposited (CVD)-grown monolayer MoS$_2$ FETs and explain electro-thermal transport using the theory of electron-phonon scattering. Both high and low field transport are found to induce perturbations in the transistor behavior which is identified after every stress cycle. Taking inference from device characteristics and using detailed analyses of the micro-Raman & Photoluminescence (PL) spectra, we eventually show that MoS$_2$ FETs undergo a unique failure mechanism wherein localized high conductance regions are formed within the MoS$_2$ channel. This phenomenon shows up as a result of field-induced material re-configuration. Subsequently, complete device failure is observed due to field induced amorphization of the channel.

## II. Results and Discussion

Field-effect transistors are fabricated on CVD monolayer MoS$_2$ for the back-gated configuration shown in figure 1(a). A scanning electron microscopic (SEM) top-view of the as-fabricated monolayer MoS$_2$ FET is shown in figure 1(b). Uniform monolayer MoS$_2$ film is verified through Raman spectra, figure 1(c).

### A. Characterization setup

In order to understand the effect of long-term electric field on device behavior, devices are stressed under different source-to-drain electric fields (E$_{SD}$) for hundreds of seconds under varying conditions like gate voltage (V$_{GS}$) and temperature (T). Current fluctuations during the stress period and low voltage device characteristics, before and after the stress cycle, are captured to identify the impact of specific stress conditions on transistor behavior. Preliminary output characteristics of a 1μm long and 10μm wide monolayer MoS$_2$ transistor, figure 2(a), show decent transistor performance with the onset of current saturation at E$_{SD}$ > 0.09 MV/cm and suggest two regimes of operation that may have more physical relevance: E$_{SD}$ ≤ 0.1 MV/cm as low-field and E$_{SD}$ ≥ 0.2 MV/cm as high-field. This is chosen as a result of current/velocity saturation and onset of breakdown observed in figure 2(b). Hence, subsequent virgin devices (VD) are stressed under these two regimes of operation at different V$_{GS}$ and T values for control over electron and phonon population in the monolayer MoS$_2$ channel respectively. In is important to note that all measurements are performed in vacuum (10$^{-4}$ torr).

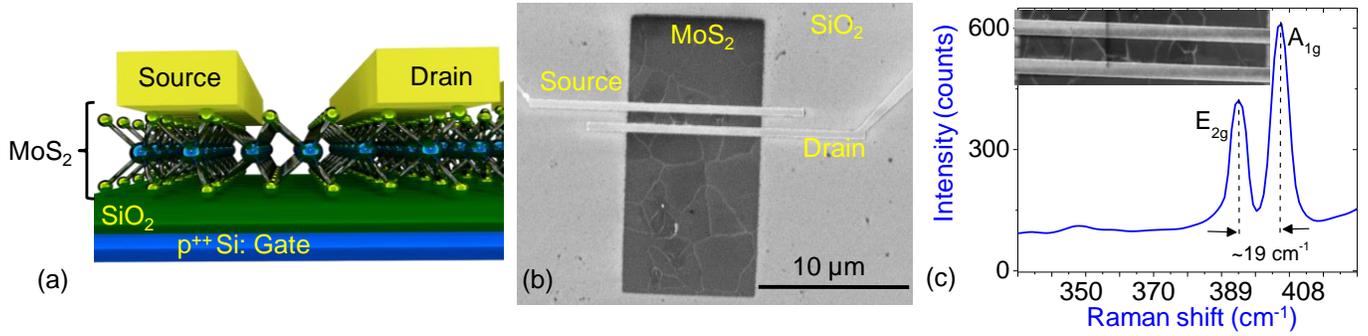

Figure 1: (a) Back-gated transistor configuration for CVD monolayer $MoS_2$ FETs used in this work for all the investigations. (b) SEM top-view of a fabricated back-gated FETs with following device dimensions: $L_{CH}$ = 1μm, W = 10μm, $T_{ox}$ = 90nm. (c) Raman spectra of the CVD $MoS_2$ film transferred on $SiO_2$ showing characteristic in-plane ($E_{2g}$) and out-of-plane ($A_{1g}$) vibrational modes with the difference between phonon wavenumbers (~19 cm$^{-1}$) implying a monolayer[29]. Inset: magnified SEM image of device channel.

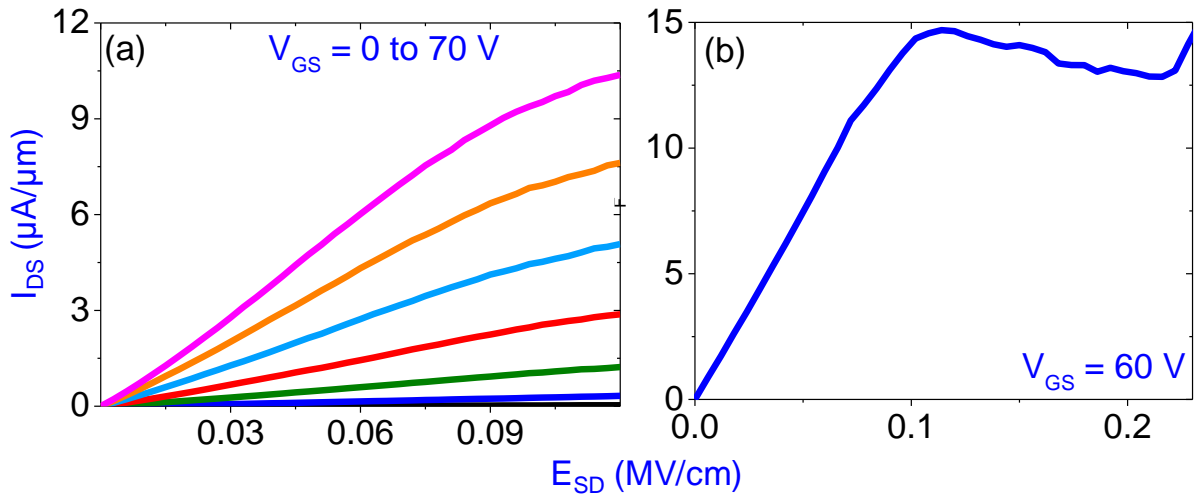

Figure 2: (a) Output characteristics of $MoS_2$ transistor exhibiting onset of current saturation at $E_{SD}$ = 0.9 MV/cm. (b) Two different regimes of operation are identified on basis of current (velocity) saturation, observed in these devices. The transistor starts to settle in saturation at $E_{SD}$ = 0.1 MV/cm and stays deep in saturation until $E_{SD}$ = 0.2 MV/cm after which breakdown occurs. For the study intended in this work, none of the devices are stressed beyond 0.2 MV/cm so that the device operates like a transistor within the safe operating regime and effect of long-term electrical stress can be captured.

**B. Steady-State Electro-Thermal Transport:**

As mentioned earlier, $MoS_2$ devices are stressed under low and high field regime: $E_{SD}$ = 0.1 MV/cm (LFR) and 0.2 MV/cm (HFR) for several seconds and stress cycles. During the stress cycles, contribution from other parameters is studied by varying $V_{GS}$ and T values. At $V_{GS}$ = 60 V (ON state), when stressed in LFR, a monotonic decrease in the channel current is observed which eventually starts to saturate after ~ 300 seconds as shown in figure 3(a). Similar variation in current at lower temperatures (150 K and 77 K) is observed however, with a small rate of decay in current. HFR operation at room temperature (300 K) results in an initial fall in current followed by sustained current through the channel, figure 3(b). It is clear from figure 3 that HFR operation results in abrupt variations in current unlike that in LFR. During low temperature HFR operation, marginal decay in current is observed for the first ~ 100 seconds after which abrupt device failure occurs resulting in extremely low current, in the range of pico-ampere, through the device. It is interesting to note that, low temperature operation in HFR introduces permanent damage to the device channel. This is also validated by the device characteristics captured after the stress cycles, as discussed in subsequent sections.

Monotonically decreasing current and abrupt fall in current are observed for LF stress and HF stress during successive stress cycles. Interestingly, the thermal equilibrium point at the end of every stress cycle shifts upwards after every stress cycle at room temperature, as shown in Fig. S1 of supplementary information (SI). In contrast, such a shift in the saturation point is not observed when devices are stressed at lower temperatures. This is shown in Fig. S2 of SI.

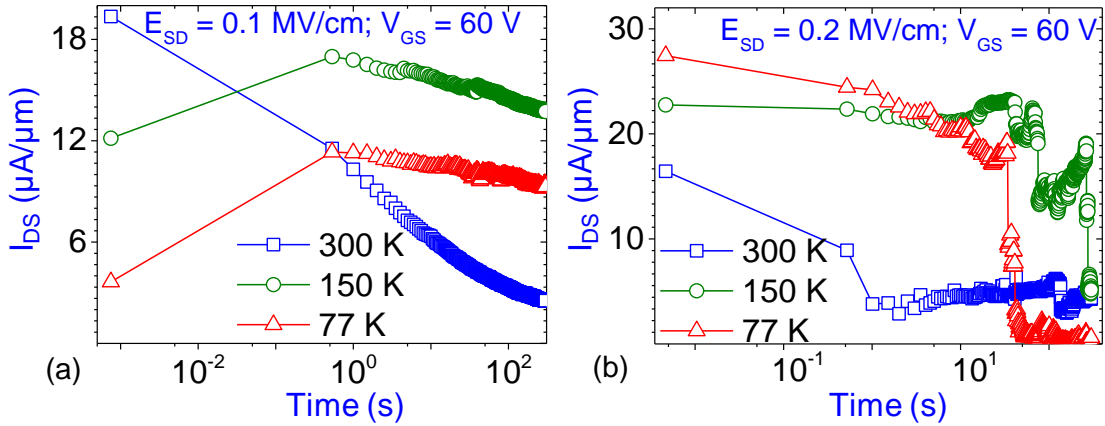

Figure 3: Temporal variation of current through MoS$_2$ channel under (a) LFR operation and (b) HFR operation at different temperatures. V$_{GS}$ is constant, maintained at 60 V during the stress period.

Device operation in LFR at different temperatures can be explained using of phonon and impurity scattering theory of electrons at high and low temperature respectively in the diffusive channel. The temperature dependent average phonon population (⟨n⟩) at an energy E (= ℏω) can be predicted using the famous Bose-Einstein distribution or Planck's distribution, i.e.

$$\langle n \rangle = \frac{1}{\exp(\frac{\hbar\omega}{kT}) - 1}$$

where ℏ = Planck's constant; k = Boltzmann constant; T = Temperature and ω = Phonon frequency are fundamental constants[30].

At any temperature, the low frequency phonon population (acoustic phonons) is much larger than the high frequency phonon population (optical phonons). Moreover, phonon population exponentially increases with temperature irrespective of phonon frequency. Therefore, events like electron phonon scattering, that involve optical phonons, have marginal probability of occurrence, especially at lower temperatures, when compared with the probability of occurrence of events like heat dissipation which essentially involves acoustic phonons. Moreover, impurity/coulomb scattering dominates at lower T whereas phonon scattering dominates at higher T[30, 31]. Keeping the above discussion in perspective, monotonic decay in current during room temperature LFR stress, as observed in fig. 3(a), is a result of electron-phonon scattering wherein an electron loses energy to the lattice and generates more phonons which manifests as heating of the channel and further scatter electrons. This continues until a thermal equilibrium is reached at ~ 100 seconds when the rate of energy lost by electron is balanced by the rate at which acoustic phonons dissipate the generated heat thereby keeping the optical phonon population constant, i.e. rate of optical phonons generated by electron-phonon scattering = rate of scattering of optical phonon into multiple acoustic phonons. Significantly small population of optical phonons and still large population of acoustic phonons at lower T and finite number of coulomb scattering sites by impurities in the channel result in a much smaller decay time constant and hence thermal equilibrium is reached much faster at lower T as shown in figure 3(a). However, initial increase in the current, figure 3(a), at lower T is attributed to acoustic phonon assisted propagation of heat dissipated by limited yet 'non-zero' number of electron-phonon scattering towards the contacts resulting in improved thermionic emission of carriers across the barrier as discussed by Abhishek et. al. for CNT devices[32].

Similarly, HFR operation at room temperature results in an initial abrupt decay of current. This is attributed to remarkable increase in the probability of electron-phonon scattering events due to high energy electrons and abundance of optical phonons at 300 K. After the initial decay in current, significant loss in the kinetic energy of electrons along with heat dissipation via acoustic phonons establish thermal equilibrium between electrons and the lattice. However, HFR operation at lower T leads to longer diffusion length because of (i) weak (but non-zero) electron-phonon scattering at low T and (ii) weak impurity scattering of high energy electrons[30]. As a result, electrons travel through the channel without losing much energy until they initiate impact ionization as shown in figure 3(b). Increase in current followed by an immediate abrupt fall is a characteristic of impact ionization process. High energy (hot) electrons bombard the lattice and generate electron-hole pairs which are then captured by the high electric field. This manifests as increased current followed by physical damage of the lattice due to disruption of bonds in the lattice by hot electrons observed as abrupt fall in current as shown in figure 3(b)[31].

*Effect of gate voltage on high-field operation (HFR)*

As discussed earlier, HFR transport in the ON state of $MoS_2$ FET results in a complete failure of the device after ~ 300 seconds of operation at lower temperatures. Partial breakdown of the device is observed at 300 K after an abrupt fall in current within few milliseconds of operation, figure 3(b). Further, electron-phonon scattering rate is a function of both electron as well as phonon population[30] which are, in principle, controlled by gate bias ($V_{GS}$) and temperature (T) respectively. Moreover, onset of device failure because of damage to the lattice due to impact ionization is determined by two quantities: high energy electron population determined by $V_{GS}$, and electron energy determined by $E_{SD}$. In order to observe the effect of electron population on the HF transport, devices are kept in HFR of operation at different gate voltages and current fluctuation is captured as a function of time, Fig. 4.

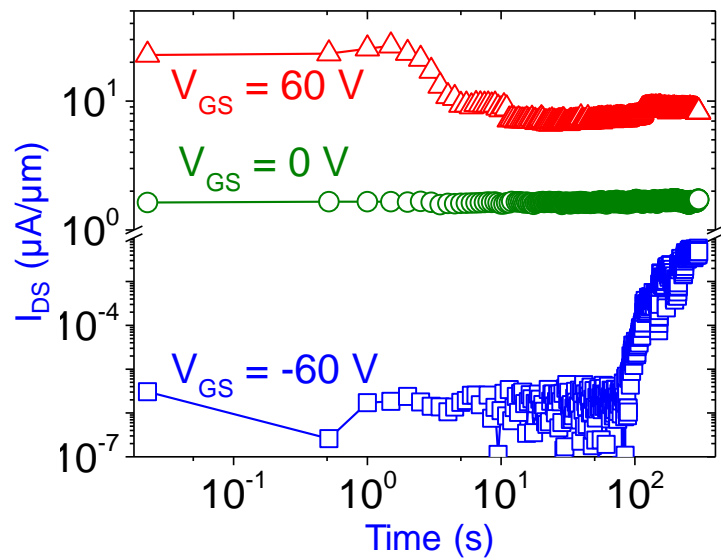

Figure 4: Effect of room temperature steady-state electrical stress (in HFR i.e. $E_{SD} = 0.2$ MV/cm) on $MoS_2$ FETs at different $V_{GS}$.

At $V_{GS} = 60$ V, as shown in figure 4, the FET is in ON state and an initial decay followed by a constant current is observed for reasons discussed earlier. However, when the carrier concentration in the channel is tuned low (via $V_{GS}$), current reduces by ~ 2 orders of magnitude, electron-phonon scattering is significantly suppressed[30]. This is reflected in the marginal variation in current at $V_{GS} = 0$ V in figure 4. Interestingly, on further reduction of carrier concentration at $V_{GS} = -60$ V, after ~100 seconds of stress, current rises abruptly by 3 orders of magnitude, figure 4. This is a unique behavior that cannot be explained using the electron-phonon scattering model at room temperature, discussed earlier. However, such a rise in current in the OFF state is possible if (i) electrons in the channel gain sufficient energy via current induced self-heating of the channel and transit from low to higher energy sub-bands in the bandstructure or (ii) significant increase in the density of states within the bandgap which facilitate current conduction through the channel in the OFF and sub-threshold region via variable range hopping at room temperature[33] or (iii) both. Among these two possible reasons, the later must introduce permanent change in the material, unlike former, which can be identified in successive stress cycles and/or transistor behavior. Moreover, current in the range of pico-ampere is not sufficient to introduce self-heating and hence transfer sufficient energy to electrons for inter-band transition. Therefore, the possibilities of the first reason for observed rise in current are minimal and a field-assisted permanent change in the material seems to be a more plausible reason. Nevertheless, to understand this behavior, impact of steady-state electrical stress on transistor behavior is studied next. Intuitively, it reveals more about the electro-thermal stress induced perturbation in $MoS_2$ and its manifestation in transistor behavior after different stress conditions, which is discussed in the next section.

**C. Impact on Transistor Behavior:**

Low voltage transfer characteristics ($V_{DS}$ = 1V) of $MoS_2$ FETs are captured for virgin (VD) and stressed devices (SD). Both HFR and LFR of operation are investigated and it is found that long-term electrical stress (both LFR and HFR) induces remarkable deviation from the original device behavior (i.e. VD), Fig. 5. Note that following are the stress conditions: $E_{SD}$ = 0.2 MV/cm (HFR) or 0.1 MV/cm (LFR) and $V_{GS}$ = 60 V unless specified.

*Impact of LF stress on transistor behavior*

After LF stress, devices exhibit higher threshold voltage ($V_T$) irrespective of temperature as shown in figure 5 (a-c). For higher temperatures (T = 300 K and 150 K), the ON state current degrades after stress. On the other hand, at T = 77 K, ON state current is found to have increased marginally (by a factor of 2) as a result of LF stress. Increase in $V_T$ is not trivial to contemplate and needs further investigations, however, decrease in ON state current at higher temperatures (figure 5(a-b)) is attributed to physical damage of the lattice caused by relatively high electron-phonon scattering. This results in significant decrease in the field-effect mobility as shown in figure 6(a-b). However, increase in ON state current after stress at 77 K (figure 5(c)) is believed to have improved because of reduction in the number impurity scattering sites that led to increased field-effect mobility, figure 6 (c).

*Impact of HF stress on transistor behavior*

HF stress, on the other hand, leads to significant increase in the OFF-state current (3 to 4 orders of magnitude) and loss of gate control. This happens especially when the device is stressed at 300 K and 150 K, as shown in figure 5(d-e). Another impact of HF stress at these temperatures is that the ON-state current degrades (2 to 3 orders of magnitude). However, HF stress at 77 K results in an overall loss of current, figure 5(f). Degraded ON state performance is a result of electron-phonon scattering induced partial physical damage to the lattice at 300 K and impact-ionization induced almost complete lattice damage at 150 K and 77 K. This is well corroborated from the observed degradation of field-effect mobility as shown in figure 6(d-f). Increase in the OFF state current and loss of gate control are attributed to electrical stress induced metal-like transport. A detailed discussion and validation of such a metal-like transport is provided in the next section.

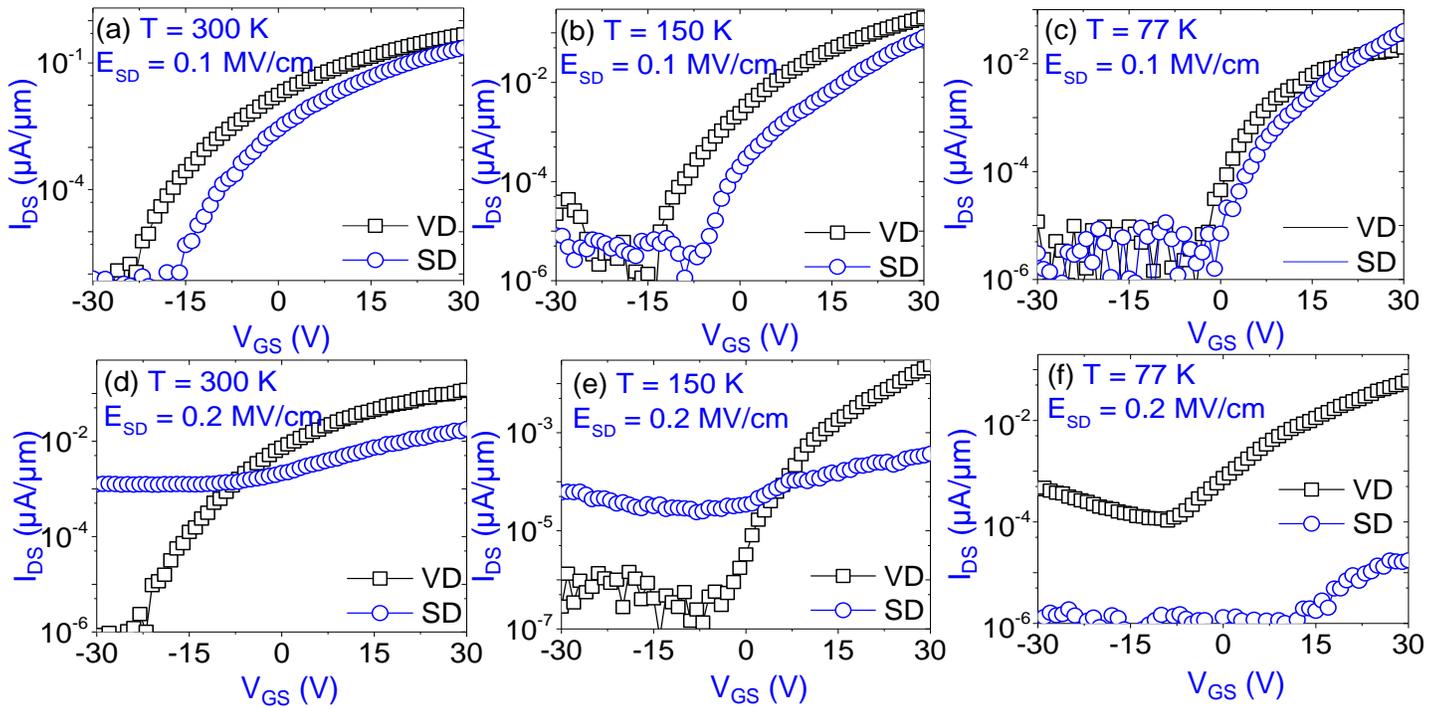

Figure 5: Transfer characteristics of unstressed/virgin (VD) and stressed (SD) FETs stressed in (a-c) LFR and (d-f) HFR of operation at different temperatures. Degraded OFF state performance, loss of gate control and reduced ON state current are the major implications of HF stress. All devices are measured at $V_{DS}$ = 1 V.

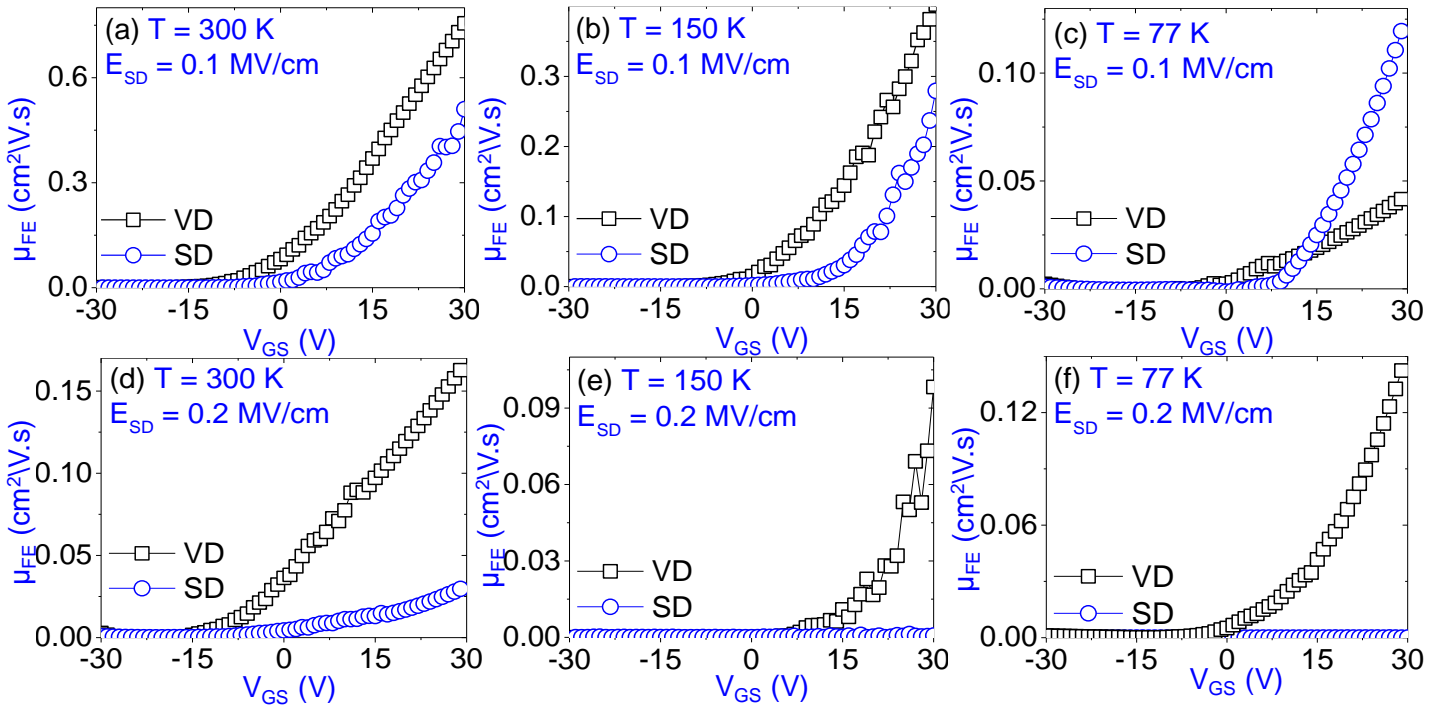

Figure 6: Field-effect mobility extracted from transfer characteristics shown in figure 5. Effect of electron-phonon scattering (a, b and d) and impact-ionization (e, f) lead to significant damage of the channel and hence reduction in mobility. (Black curve: VD & Blue curve: SD)

*Effect of electrical stress at lower gate voltage*

Interestingly, devices stressed at lower gate voltages, shown in figure 4, also exhibit high OFF state current and poor gate control as shown in figure 7. However, in this case the ON state current increased after the stress cycle. Whereas, as discussed earlier, a stress cycle of $E_{SD}$ = 0.2 MV/cm at $V_{GS}$ = 60 V leads to abrupt current decay with time and correspondingly results in higher OFF state current, poorer gate control and smaller ON state current. It is important to note that HF stress results in higher OFF state current and degraded gate control of the transistor channel irrespective of the gate voltage during stress cycle. However, ON state current depends on the gate voltage which essentially determines the number of electron-phonon scattering events at room temperature. It appears that higher OFF state current and poor gate control are fallout of a certain lateral field-dependent phenomenon that results in a highly conductive (metal-like) channel when the device is in the OFF state (figure 4-d and 7). Along with electron-phonon scattering, the same phenomenon determines ON state current i.e. higher (lower) gate voltage during stress degrades (improves) ON state current as shown in figure 4-d (figure 7).

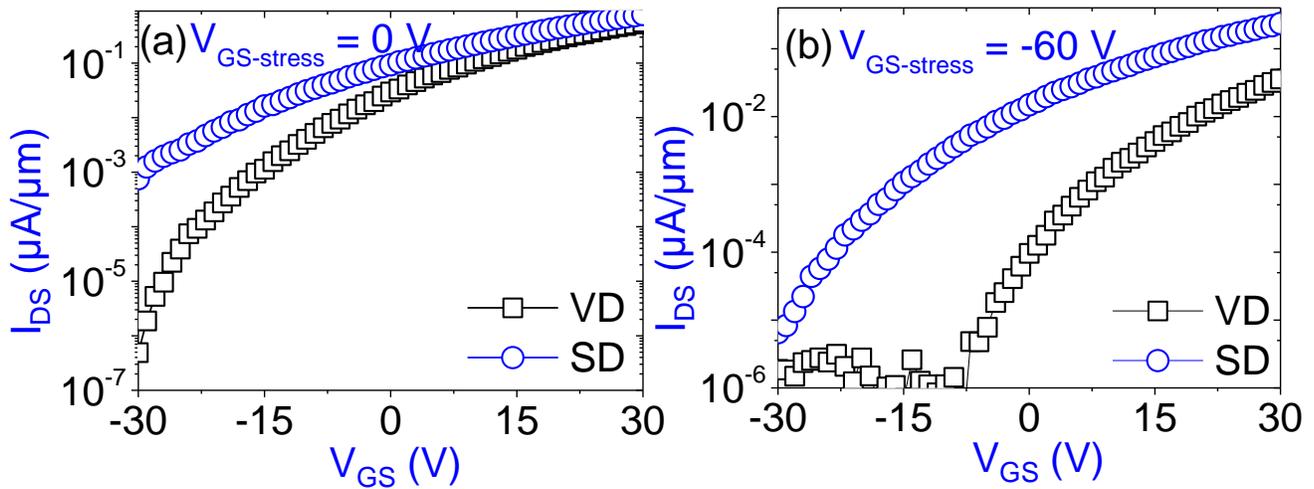

Figure 7: Transfer characteristics of unstressed (VD) and stressed (SD) FETs. SD is stressed at $E_{SD}$ = 0.2 MV/cm and $V_{GS}$ = (a) 0 V and (b) -60 V.

In order to validate stress-enhanced conductance of the channel, 2T terminal channel current ($I_{DS}$ at gate open condition-figure 8-a) of VDs are measured and compared with that of SDs stressed in the HFR. As shown in figure 8 (b & d), devices stressed at $V_{GS}$ = 60 V and T = 300 K (to be identified as D-HF-60-300 throughout the text henceforth) and 77 K (to be identified as D-HF-60-77 throughout the text henceforth) exhibit lower conductance after stress. However, stress at 150 K (to be identified as D-HF-60-150) (figure 8-c) results in increased channel conductance. Similarly, when devices are stressed at $V_{GS}$ = 0 and -60 V, the device conductance is found to have increased as shown in figure 8(e-f). Interestingly, upon successive stress cycles, higher conductance in D-HF-60-300 whereas lower conductance in D-HF-60-150 are observed, figure 8-b and 8-c (green curve) respectively. Increase in conductance after different stress conditions and cycles can be correlated with poor gate control and higher OFF state current observed in the transfer characteristics earlier. These observations clearly imply that the channel certainly becomes more conductive upon stress under specific conditions and extent to which the channel conductance increases upon stress depends on the gate voltage applied during the stress cycle; for example 1(5) order(s) of magnitude increase in conductance at $V_{GS\text{-stress}}$ = 0 V (-60 V). This is attributed to interplay between two mechanisms, (i) electron-phonon scattering resulting in reduced carrier mobility in the channel and (ii) speculated material re-configuration resulting in higher conductance. Under other stress conditions like HF stress at 77 K and subsequent stress cycle at 150 K, channel conductance decreases significantly. This is attributed to impact ionization-induced wide spatial distribution of physical damage of the material under these stress conditions as observed earlier in current versus time curve in figure 4.

Increase in the conductance of the device along with poor gate control and negative shift in $V_T$ implies increase in free-electron concentration in the channel that prevents gate electrode to effectively tune the channel conductance. Moreover, increase in the density of mid-gap states may cause higher sub-threshold and OFF state current by increasing the probability of variable range hopping of electrons between mid-gap states at room temperature[33]. In order to validate the presence of electrical stress induced increase in the electron concentration, Raman and PL spectroscopy of stressed $MoS_2$ is performed which is discussed in the next section.

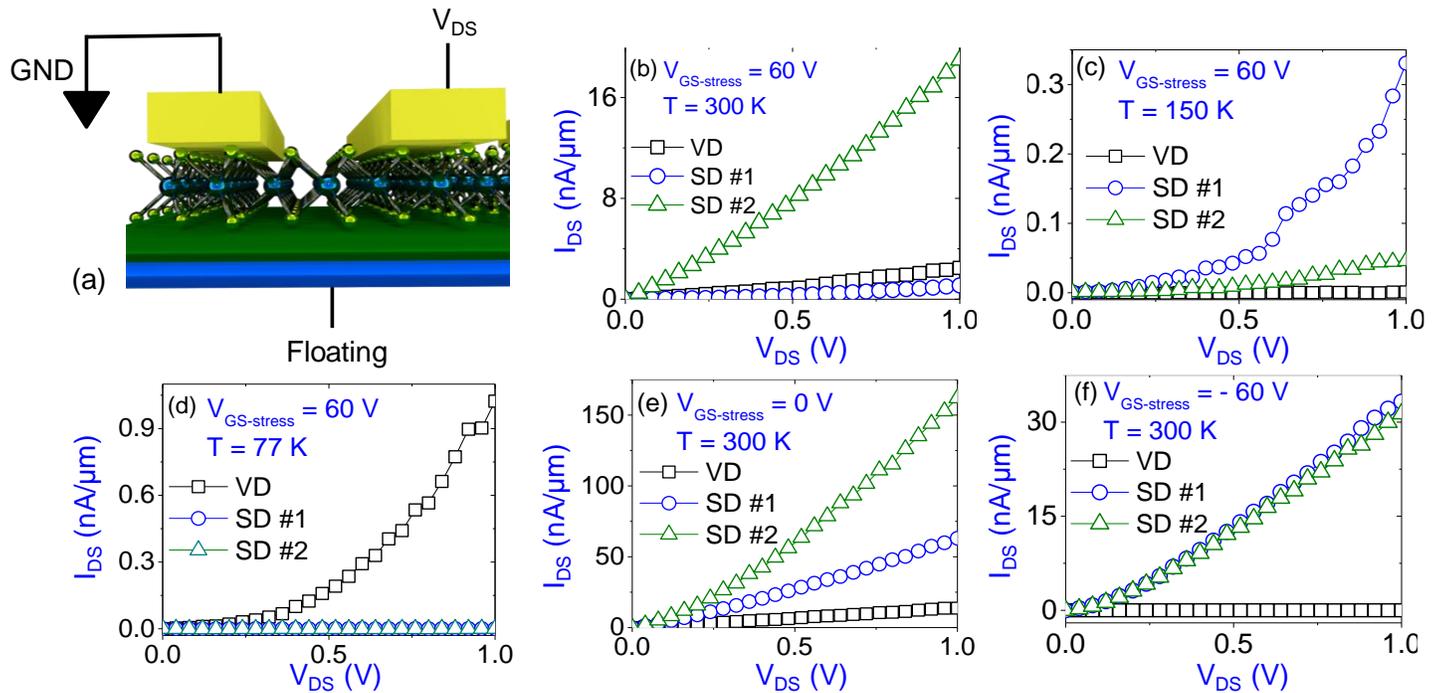

Figure 8: (a) Bias condition for measuring the 2-terminal current in the channel i.e. when the transistor behaves like a 2-terminal resistor. 2-terminal current of unstressed (VD) and stressed (SD) devices in the HFR at (b) 300 K, (c) 150 K and (d) 77 K is measured as a function of $V_{DS}$. In order to observe the effect of gate voltage variation, 2-terminal current is measured after before and after stressing the device under two different conditions: (e) HFR at $V_{GS-stress}$ = 0 V and (f) at $V_{GS-stress}$ = -60 V both at T = 300 K. Here SD #1 (SD #2) implies that the SD device is stressed once (twice).

## D. Electrical Stress Induced Material Degradation

Raman spectra of different locations across the channel are captured and compared with that of unstressed $MoS_2$. As shown in figure 9(a), the out-of-plane mode, $A_{1g}$, is found red shifted at multiple positions of $MoS_2$ channel stressed in the HFR at 300 K. A red shift implies reduced energy of that mode which is essentially the out-of-plane vibrations of the S atoms in $MoS_2$. This has been related to increased electron population in $MoS_2$ [34]. Further, $MoS_2$ stressed at 150 K also exhibits signs of increased electron population with a marginal red shift of the $A_{1g}$ mode, figure 9(b). These results strongly validate that, as observed earlier in device characteristics, electrical stress induces localized regions with excess electrons that can potentially lead to increased OFF state current, poor gate control and negative shift in $V_T$. However, for $MoS_2$ stressed at 77 K, it is observed that the characteristic peaks are not prominent. Moreover, non-zero intensity peaks of the Raman characteristic modes are distributed throughout the range of Raman shift shown in figure 9(c). This is attributed to amorphization of the $MoS_2$ and implies physical breakdown of the channel has not occurred. This is in-agreement with (i) earlier observations of device failure after HF stress at 77 K (figure 5-f) and (ii) scanning electron microscope (SEM) top view shown in figure 10(a) where a physical break in the channel is not observed. Another important evidence of amorphization is the Kelvin Probe Force Microscopy (KPFM) image of the device stressed in HF at 77 K shown in figure 10(b). KPFM is used to measure surface potential where grain boundaries, mostly S defects, show up as low potential regions. Another low potential region in a typical device scan is the metal electrode. In an arbitrary CVD $MoS_2$ device, grain boundaries run randomly in and out of the channel without a specified orientation. Typically, if the grains are moderate in size and channel length of the device is large enough, multiple grain boundaries are found to be present within the channel, as shown in Fig. S3(a) of supplementary information (SI). However, in a device stressed in HFR at 77 K, grain boundaries are not observed anywhere within the channel which is attributed to complete amorphization of material resulting in multiple orientations of the crystal at molecular level, difficult to resolve in KPFM. This is also observed in the KPFM scan of a device stressed at 150 K, shown in fig. S3(c) of SI. Another possibility, complete crystallization of MoS$_2$ (also without grain boundaries), however, is not obvious because of disagreement with device results in figure 3 and 5(f). KPFM scan of devices stressed in HFR at V$_{GS}$ = 60 V and 0 V are shown in Fig. S3 (b & d) of SI. Unusual low potential (dark) regions are observed in the channel between source and drain contacts. These regions are believed to be the source of metal-like behavior observed in transistor behavior where gate control over channel current is lost after the stress. Details of the KPFM measurement setup are discussed in section 3 and S4 of SI.

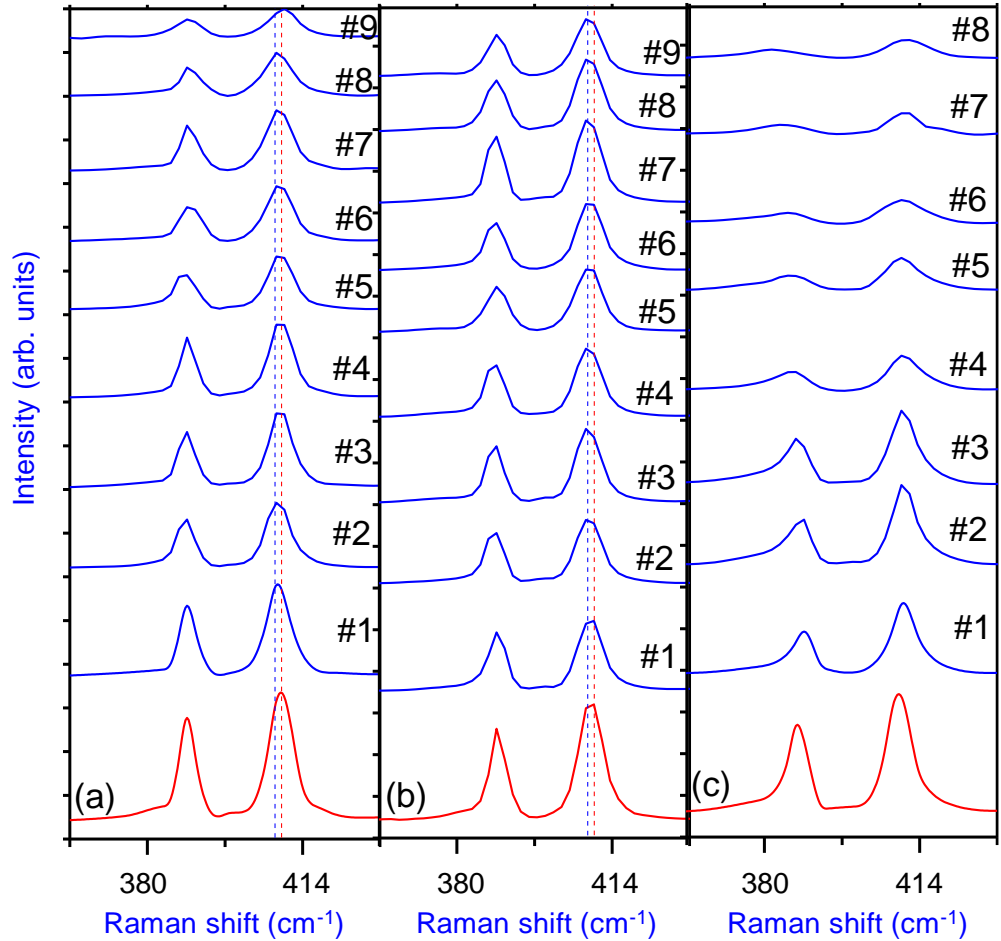

Figure 9: Raman spectra of stressed (under different stress conditions: HF stress with V$_{GS}$ = 60 V at T = (a) 300 K, (b) 150 K and (c) 77K) MoS$_2$ channel at 10 different spots (blue curves) along the width of the device, W = 10 µm, compared with that of an unstressed MoS$_2$ (red curve) region. Note that Raman laser spot size is 1 µm in diameter.

Besides transistor characteristics and micro-Raman spectra, PL spectra of stressed and unstressed MoS$_2$ channel is also found to strongly validate the presence of electrical stress-induced formation of highly conductive localized regions within the channel. Monolayer MoS$_2$ is known to exhibit strong PL due to its direct bandgap and strong excitonic features encouraged by its atomically thin structure where electrons and holes are strongly constrained within the monolayer resulting in, strong coulombic force. This facilitates sustained formation of various excitonic species. As shown in figure 11(a), PL spectra of pristine monolayer MoS$_2$ has three characteristic peaks: (i) B exciton, (ii) neutral A$^0$ exciton and (iii) negative A$^-$ trion, which are obtained after deconvoluting the effective PL signal obtained from the CCD detector. These peaks correspond to the various photon emission events that occur after incident photons with certain energy create electron-hole pairs which subsequently recombine. While B and A$^0$ excitons are single electron-single hole pairs that emit characteristic photons due to conduction band (CB) to light hole and heavy hole valence bands (LHVB and HHVB) respectively, A$^-$ trion is a two electron-single hole pair which emits a relatively low energy photon on recombination from CB to VB. PL spectra of stressed and unstressed regions are captured at different positions along the width of the channel and deconvoluted to identify different peaks.

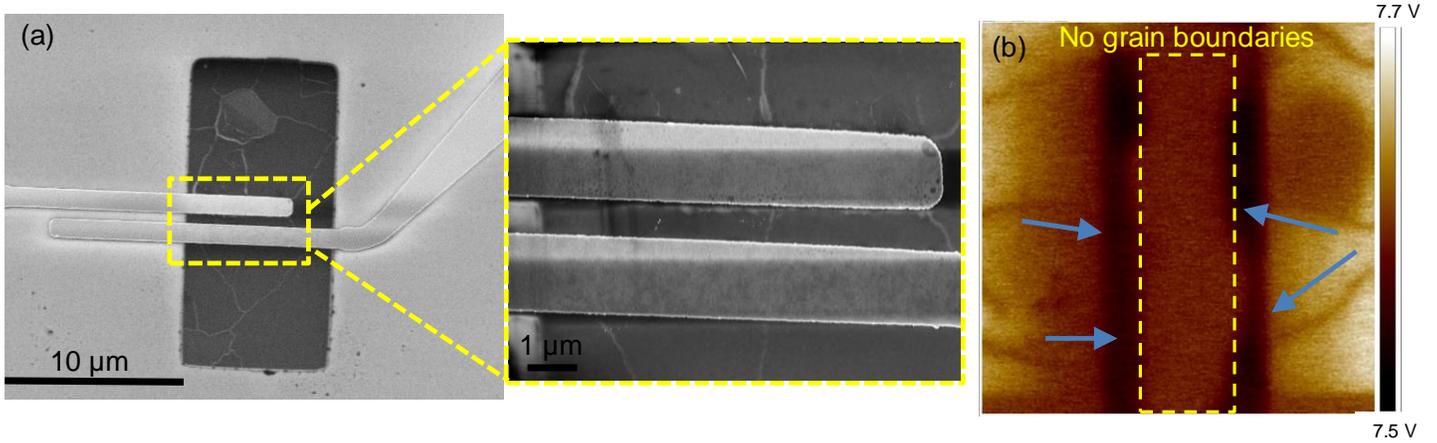

Figure 10: (a) SEM top view image of D-HF-60-77 and inset showing continuous channel. This implies that impact ionization did not result in physical discontinuities in the channel. (b) KPFM image of the same device implying that the stressed channel does not have grain boundaries that are otherwise expected because of their orientation outside the channel depicted by blue arrows. Dark regions in the image are metal electrodes, grain boundaries between two $MoS_2$ grains and other impurities that may have low potential. Bright region is $MoS_2$ grains that have higher potential than metal and grain boundaries.

In Fig. 11(b), it is observed that the intensity ratio of trion peak to neutral exciton peak significantly increases at multiple locations of the device stressed in the HFR at 300 K. For devices stressed at 150 K and 77 K, this ratio remains closer to that in a virgin device, as denoted by dotted horizontal line in Fig. 11(b). Remarkable increase in this ratio implies that the number of photons emitted due to negative trion is larger than that emitted from annihilation of a neutral exciton. This suggests that there are more number of trion species formed in the channel stressed at 300 K than neutral exciton species. This is attributed to increase in the free electron concentration[34] and hence validates that, devices stressed at 300 K have multiple localized regions within the channel with higher conductance. Moreover, regions (from position #7 to #9) within the channel of device stressed at 77 K fail to exhibit any of the characteristic peaks. This is attributed to significant damage to the lattice and is also validated by full-width half-maximum (FWHM) values shown in Fig. 11(c-e). FWHM of PL characteristic peaks, especially $A^0$ and $A^-$, increases for devices stressed at 150 K and 77 K, Fig. 11(d,e). While for devices stressed at 150 K, a decent increase in FWHM is observed, for devices stressed at 77 K, FWHM is found to have increased by ~ 2× for $A^0$ peak and ~ 3× for the $A^-$ peak along with huge deviation in FWHM from one channel region to another, as depicted by the error bar in Fig. 11(c-e). Typically, FWHM of a PL peak is related to the density of defects in the region of scan. A roughly defined (broad) PL peak (large FWHM) denotes large density of defects as compared to a sharp (small FWHM) PL peak. Therefore, Fig. 11(c-e) validates the fact that HF stress at lower temperature induces significant damage to the lattice, amorphized channel as discussed above, which eventually results in marginal current through the device.

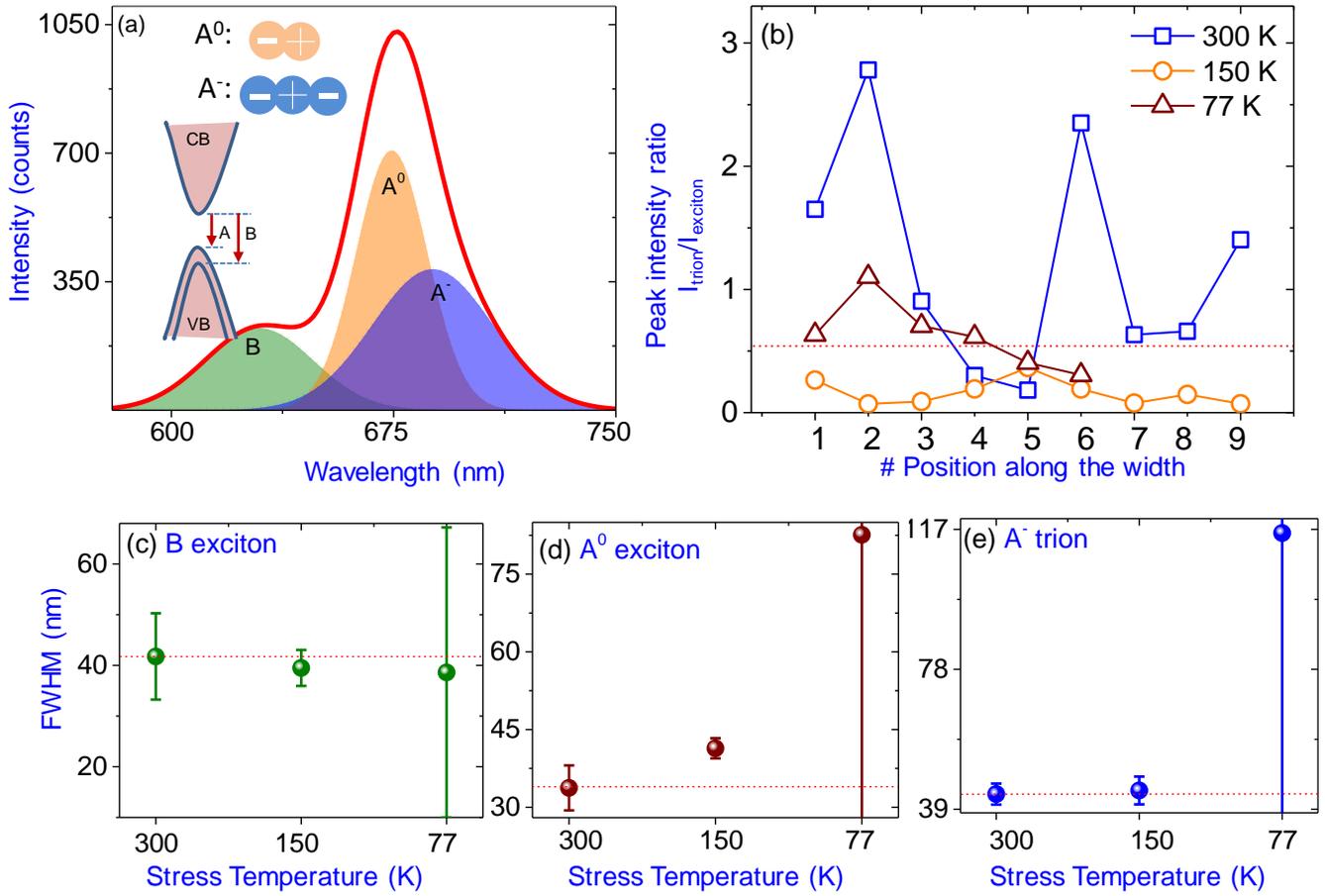

Figure 11: (a) PL spectra of monolayer $MoS_2$ consisting three characteristic peaks corresponding to three different direct energy transitions: B exciton, $A^0$ neutral exciton and $A^-$ trion. (b) Intensity ratio of trion and neutral exciton peaks at different locations along the width of the channel of devices stressed at different temperatures: 300 K, 150 K and 77 K. The dotted horizontal line denotes intensity ratio in a virgin monolayer $MoS_2$ channel. Increase in the trion to exciton peak ratio for device stressed at T = 300 K implies that the electron concentration is higher than that in a virgin channel. For devices stressed at 77 K, there is no PL signal beyond position 6 which is attributed to physical damage led discontinuity in the lattice. Physical damage or increases in defect concentration can be qualitatively understood by increase in the full-width half maximum (FWHM) of various peaks in the PL spectra. Average FWHM (averaged over different locations along the width of the stressed channel) of (c) B exciton peak, (d) neutral exciton peak and (e) negative trion peak as a function of temperature during electrical stress implies that high voltage electrical stress at lower temperature induces significant damage to the lattice. The dotted red line is the FWHM of different peaks in a virgin $MoS_2$ channel. Peak intensity ration and comparison of FWHM for different peaks are extracted from PL spectra for these devices which is shown in Fig. S5 of SI.

While it is clearly identified that CVD-grown monolayer $MoS_2$ undergoes amorphization before complete device failure, the origin of enhanced conductance upon electrical stress is not clear, even after enough validation. Interestingly, this behavior is similar to the one observed by Sangwan et. al. where $MoS_2$ exhibits resistance switching at SET voltages as high as 30 V[36]. This is attributed to migration of Sulfur vacancies within the channel under the effect of external electric field. Such a behavior has been found to be assisted by grain boundaries present in CVD monolayer $MoS_2$. S vacancy migration under electric field is a potential justification of the observed localized region with excess electrons in an appropriately stressed $MoS_2$ channel. These localized regions could possibly have accumulated S vacancies and once aligned continuously between source and drain, they result in loss of gate control, negative shift in $V_T$ and higher OFF state current in $MoS_2$ transistor. The fact that S vacancies in $MoS_2$ manifest as mid-gap states ithin the bandgap, justifies that long-term electrical stress increases the density of these mid-gap states, as speculated earlier in sections 3 and 4, by re-distributing the S vacancies in the channel. Moreover, when stressed at lower $V_{GS-stress}$, the channel also exhibits high ON state current due to two reasons: (i) formation of localized high conductive regions, as discussed earlier and (ii) significantly lower electron-phonon scattering induced lattice damage.

## III. Conclusion

CVD monolayer MoS$_2$ FETs are operated under low and high field regimes to unveil electro-thermal transport in MoS$_2$. Low field transport in MoS$_2$ can be well explained using electron phonon scattering model at different temperatures and gate field. High field transport triggers impact ionization led device failure which is relatively easier to accomplish at lower temperatures. Moreover, it turns out that device failure occurs due to amorphization of the channel due to high field steady state operation. Before complete failure, both low and high field steady state device operation at room temperature lead to perturbation in transistor behavior wherein the OFF-state performance and gate control degrade which imply metal-like behavior. This unconventional device behavior depends on stress conditions like gate bias, stress voltage and number of stress cycles and is validated by 2-terminal current measurement and micro-Raman & Photoluminescence (PL) spectra. While 2-terminal current increased after high field stress, red shift in the A$_{1g}$ mode and enhanced negative trion peaks of different regions within the channel clearly imply the presence of localized regions with higher free-electron concentration. This proves that the performance of MoS$_2$ transistor degrades under long-term electrical stress by inducing highly conductive regions within the channel and a subsequent loss of gate control. Such a failure mechanism is unusual and understanding its origin is critical for determining a regime for safe operation of MoS$_2$ transistors. Moreover, such a mechanism limits the terminal voltages for which transistors exhibit acceptable performance. These findings also suggest that voltage scaling employed for Silicon transistors may not be the same for MoS$_2$ and one needs to be more careful in determining these limits because unlike in Silicon, a prolonged operation of an MoS$_2$ transistor leads to significant increase in the channel current which can have detrimental effects on the overall circuit performance.

## IV. Methods

*Device fabrication*: Monolayer CVD MoS$_2$ film grown on Sapphire is transferred onto a 90 nm thick thermally grown SiO$_2$ on Si substrate and cleaned in acetone and IPA to ensure minimum organic residue on the film. The sample is patterned (using electron beam lithography) in order to define regions for device fabrication. By using Oxygen plasma, unwanted MoS$_2$ is etched out inside a reactive ion etching (RIE) chamber followed by patterning (via electron beam lithography) of regions for source/drain electrode (Ni/Au: 10/50 nm) deposition inside an electron beam evaporator. Subsequent metal lift-off and vacuum anneal for 10 minutes at 250 °C ensure removal of unnecessary metal from the sample and formation of high-quality metal-MoS$_2$ interface.

*Kelvin Probe Force Microscopy*: All KPFM measurements are performed using Bruker Dimension Icon AFM setup under ambient conditions. The AFM tips used here are Pt/Ir coated Silicon probes-SCM-PIT-V2 with a radius of ~ 25 nm. Scanning a surface to capture potential profile using this AFM setup is a two-pass method in which the tip first scans the surface via line scans in tapping mode followed by a second scan (second pass) during which the tip is lifted at a height (15 nm in this case) and a DC bias is applied to the tip (7 V for these measurements) along with a small AC signal while scanning alone the same line with a scan rate of 0.7 Hz. During the first pass, the cantilever is made to oscillate close the its resonant frequency.

**Data Availability**

The data to support findings of this work are available from the corresponding author upon reasonable request.

**Acknowledgment**

Authors would like to thank NNetRA program of MeitY, DST and MHRD, Govt. of India, for supporting this work.

## Conflicts of interest

There are no conflicts to declare.

## Author Contribution

Ansh and GS fabricated all the devices. Ansh performed electrical measurements along with SEM imaging and KPFM scans. Ansh, MS and JK analyzed the data. Ansh and MS wrote the paper.